\documentclass{aa}  
\usepackage{graphicx}
\usepackage[T1]{fontenc}
\usepackage{amsmath}
\usepackage{graphicx}
\usepackage{txfonts}
\usepackage{natbib} 
\bibpunct{(}{)}{;}{a}{}{,} 

\newcommand{\ve}[1]{{\bf #1}}

\begin{document}

\title{Magnetic coupling of planets and small bodies with a pulsar wind.}


   \author{F. Mottez \inst{1}            
          \and
          J. Heyvaerts \inst{2,1}\fnmsep
          }


   \institute{LUTH, Observatoire de Paris, CNRS, Unviersit\'e Paris Diderot,
              5 place Jules Janssen, 92190 Meudon, France.\\
              \email{fabrice.mottez@obspm.fr}
         \and
             Observatoire Astronomique, Universit\'e de Strasbourg,
		11, rue de l'Universit\'e, 67000 Strasbourg, France.\\
             \email{jean.heyvaerts@astro.unistra.fr}}


 
  \abstract{}{We investigate the electromagnetic interaction of a relativistic stellar wind with a planet or a smaller body
  in orbit around the star. 
  This may be relevant to objects orbiting a pulsar, such as PSR B1257+12 and PSR B1620-26 
  that are expected to hold a planetary system, or to pulsars with suspected asteroids or comets.\\ }
  {We extend the theory of Alfv\'en wings to relativistic winds. 
  \\}
  {When the wind is relativistic albeit slower than the total Alfv\'en speed, a system of electric currents carried by 
  a stationary Alfv\'enic structure is driven by the planet or by its surroundings. 
  { {For an Earth-like planet around a "standard" one second pulsar, the associated current can reach the same magnitude as the Goldreich-Julian current that powers the pulsar's magnetosphere.}} 
    }{}

   \titlerunning{Alfv\'en wings}
   \authorrunning{Mottez and Heyvaerts}
   \maketitle
%


\section{Introduction} \label{sec_intro}

Precise pulsar timing measurents proved that the pulsars
PSR B1257+12 and PSR B1620-26, host planets, at distances of order of an astronomical unit 
(\citet{Wolszczan_1992}, \citet{Thorsett_1993}).
Moreover, accretion discs are expected 
to form at some phase of the evolution of neutron stars in a binary system,
possibly giving birth to second generation planets. 
Small bodies, such as planetoids, asteroids or comets may also orbit pulsars  
and occasionally fall on them.

Circum-pulsar objects 
move in the centrifugally driven relativistic pulsar wind.
The angular velocity $\Omega_*$
of a rotating neutron star 
typically is larger than 10 rad.s$^{-1}$. The star
behaves like an antenna \citep{Deutsch_1955} 
emitting by magnetic dipole radiation a power
which causes it to lose rotational energy at a rate
\begin{equation} \label{eq_puissance_dipole_rotation}
\dot E_{rot} =- M_I \Omega_* \dot \Omega_* =4 \pi^2 M_I \dot P / P^3,
\end{equation}
where $M_I \approx (2/5)M R^2$ 
is the neutron star's moment of inertia \citep{Lyne_1998}.
This power can be compared to the 
orbital energy of a pulsar planet.
Consider, for example, 
the case of PSR B1257+12 and its planet "a". Using the data in Tables \ref{table_input_pulsars} 
and \ref{table_input_planetes}, we can estimate the moment of inertia  to be
$M_I = 1.\times 10^{38}$ kg.m$^{2}$, 
the star's rotational energy loss to be $\dot E_{rot} = -2.\times 10^{27}$ W, 
and the planet's orbital energy to be $E_G = {G M_* M_P}/{2 a }  = 4.  \times 10^{32}$  J, 
where $a$ stands for the planet's 
semi-major
axis . 
The planet and its environment
intercept a fraction larger than or of order of $\pi R_P^2/(4\pi a^2)$ of this power. This captured power is used to heat the planet, 
to generate the current system described below and to work on the planet's motion 
as described in the accompanying paper \citep{Mottez_2011_AWO}, hereafter (MH2). 
The planet's radius can be infered from its mass by
assuming a terrestrial density (5000 kg m$^{-3}$).
The intercepted power ${\dot{E}}_{capt}$ is~:
\begin{equation}
{\dot{E}}_{capt} = {\dot E_{rot}} \left(\frac{R_P}{2a}\right)^2.   
\end{equation}
For planet "a" of PSR B1257+12, we find,  ${\dot{E}}_{capt} =$   
1.74 10$^{18}$ Watts. If a substantial part of this power goes into
performing work on the planet's motion, an orbital evolution time scale of about 8. 10$^6$ years may be expected. 
This is a short time scale by astronomical standards. It scales with the mass $M_P$ of the planet as $M_P^{1/3}$. 

{{In this paper, we examine the  interaction of these planets with the magnetized wind of their pulsar and discuss
in (MH2) the effect of this interaction on the long term evolution of their orbital elements.}}

{{The energy flux carried by the wind of an ordinary star, such as the Sun,
is small enough to have but negligible}} effect on the orbits of its planets. 
Stellar winds {{usually are asymptotically super-Alfv\'enic, by which we mean that they eventually become faster 
than the total Alfv\'en speed, associated with the modulus of the magnetic field.
As a result, the planets are  protected from a direct contact}} with
the wind by a bow shock. 

Pulsar winds are much different. In a first approximation, {{oblique rotator pulsars may be regarded as}}
magnetic dipoles rotating at high angular velocity in vacuo, which causes a low frequency
and large amplitude electromagnetic wave to be emitted, {{the wave character of which reveals itself 
in the wave zone}}, beyond the light cylinder, of radius $r=c/\Omega_*$.
In this zone, magnetic field lines become spiral-shaped and the azimuthal component
of the magnetic field decreases as $B_\phi \sim r^{-1}$
while its radial component 
decreases as 
$B_r \sim r^{-2}$. The magnetic field becomes mostly azimuthal at large distances 
and at the equatorial latitudes were the planets are expected to be found.  {{Aligned rotators 
essentially are rotating unipolar inductors which generate, beyond the light cylinder, an highly 
relativistic MHD wind in which the magnetic field also becomes predominantly
azimuthal at large distances from the rotation axis, the field components 
$B_\phi$ and $B_r$
varying with distance essentially as indicated above, at least near the equatorial plane and
when the flow is close to being radial. 
The power emitted by such objects is also of order of $\dot E_{rot}$, given by Eq. (\ref{eq_puissance_dipole_rotation}), because 
the Poynting flux emitted through the light cylinder is comparable for wind or wave emission. 
We only consider aligned rotators in this paper. Their analysis is simplified by the fact
that the magnetic field of the wind observed in the
planet's frame is close to being time-independent.
The origin, acceleration and structure of pulsar winds are not fully understood. 
A number of models have been proposed in the literature \citep{Michel_1969,Henriksen_1971,Contopoulos_1999,Michel_2005,Bucciantini_2006}.
In spite of their diversity, they all converge on the fact that the wind is dominated by the Poynting flux, {{although
observations indicate that they turn into matter-dominated high energy flows at large distances
\citep{Kirk_2009}. At distances of order of an astronomical unit,
pulsar winds are expected to be still Poynting-flux-dominated.}} 
{ {This means that the electromagnetic  energy density $\sim B_0^2/\mu_0$, 
$\mu_0$ being the magnetic permeability of vacuum,  is much larger than the 
plasma energy density $\gamma_0 \rho_0 c^2$, $\gamma_0$ being its Lorentz factor 
and $\rho_0$  the rest mass density of this supposedly cold wind. 
These quantities refer to the observer's frame.}}
In such circumstances, Alfv\'enic perturbations
propagate at a phase velocity close to the speed of light (equation (\ref{eq_va})). 
Therefore, in spite of being highly relativistic, the wind flow 
may nevertheless be sub-Alfv\'enic, in the sense defined above.
A planet in a Poynting-flux-dominated wind may remain unscreened from the wind 
by a bow shock and thus enter in direct contact with it. 

The interaction of a planet with a sub-Alfv\'enic plasma flow has been 
considered for moderately magnetized non-relativistic flows in connection with the
interaction of the satellite Io with the
plasma and magnetic field present in Jupiter's magnetospheric environment.
This interaction is driven by the inductive electromotive field which results from the motion
of the satellite across Jupiter's corotational magnetic field and plasma flow.
The satellite acts as a (uniformly moving) unipolar inductor.
\citet{Neubauer_1980} derived a nonlinear theory 
of this interaction.
The Alfv\'en wing connecting Io and Jupiter is the only explored case of such a structure
in the universe up to now, and it has been the object of recent studies concerning 
its overall structure \citep{Chust_2005,Hess_2010}, the possibility of particle 
acceleration \citep{mottez_2007_a,mottez_2009_a} and its consequences on the radio emissions 
\citep{Queinnec_1998,mottez_2007_b,mottez_2009_b}. 
In the present paper, we develop a theory that generalizes some of Neubauer's results 
to the case of highly magnetized ($B_0^2 >> \mu_0 \rho_0 \gamma_0 c^2$) and relativistic plasma flows 
with Lorentz factors $\gamma_0 >> 1$, 
when a planet immersed in the magnetized pulsar wind
acts as a unipolar inductor
and generates two stationary Alfv\'enic structures that emerge from the planet and extend far in the wind.



\section{A unipolar inductor in the pulsar wind} \label{sec_unipolar_inductor}
{Let us consider a planet orbiting a pulsar in the relativistic flow of the emitted wind. 
Different reference frames can be involved in the description of the fluid motion. 
The frame where
the neutron star is at rest is the observer's frame. We denote it by $R_O$. 
The planet velocity being small compared to the wind velocity,
we may consider the planet to be at rest with respect to the neutron star, except when
discussing the planet's motion.  
The reference frame $R_O$ can then also be regarded as being the planet's rest frame. 
Quantities observed in this frame are denoted by letters without any
superscript, such as $\rho$ or ${\mathbf{v} }$. 
The unperturbed wind's instantaneous rest frame in the vicinity of the planet 
is the "wind's frame" $R_W$. Quantities observed in this frame are denoted by letters with a
prime superscript, such as $\rho'$. An index $0$ refers to quantities associated with the unperturbed wind.
The unperturbed wind velocity in $R_O$ is ${\mathbf{v}}_0$, its associated Lorentz factor is
$\gamma_0$, the unperturbed magnetic field is ${\mathbf{B}}_0$ and the wind's density is $\rho_0$. 
The wind plasma being supposedly cold, this mass density reduces to the (apparent) rest mass density of particles.
In the wind frame $R_W$, the unperturbed wind density is its proper rest mass density $\rho'_0$ 
and its magnetic field is ${\mathbf{B}}'_0$. They both differ from  $\rho_0$ and ${\mathbf{B}}_0$.
In the presence of a perturbation, the instantaneous rest frame of the fluid is not $R_W$ but
another rest frame, $R_{F}$. Quantities observed in this instantaneous 
rest frame are indicated by a subscript $F$.}

{{The perturbation generated by the planet in this flow
is time-dependent in the wind's frame. Since in the tenuous and highly
magnetized pulsar wind the formal Alfv\'en velocity $c_A$ may exceed the speed of light,
}}
the derivation of the propagation velocity $V'_A$ of Alfv\'enic perturbations must take into 
account the displacement current. In the wind's rest frame~:
\begin{equation} \label{eq_va}
V_A^{'-2}=c^{-2}+c_A^{'-2}=c^{-2}+ \mu_0 \rho'_0 /{B'_0}^2.
\end{equation}  
If the flow is faster than the Alfv\'en velocity $V_A$, the planet is 
preceded by a shock wave that defines a confined 
area where the flow is strongly modified. 
But if the flow is sub-Alfv\'enic, the planet is directly in contact with the wind.
In the {{instantaneous}} rest frame of the {{wind plasma
we expect, following the MHD approximation, the electric field to vanish.
In the planet's rest frame}} 
an electromotive field $\ve{E} = -\ve{v} \times \ve{B}$ is generated,
where $\ve{B}$ and $\ve{v}$ are {{the magnetic field and fluid velocity in the planet's frame}}.

{{Using a model pulsar wind, it is possible to make an estimate 
of the electric field.
One of the simplest such models was developed by \citet{Michel_1969}.
It includes}}
the magnetic field and the star's rotation, wind particles masses, 
but  neglects the dipole inclination and the gravitation {{and assumes the wind flow to be radial}}. 
The unperturbed magnetic field ${\mathbf{B}}_0$ only has a radial poloidal component $B_0^r$ and an azimuthal toroidal component
$B_0^\phi$. 
{The neglect of the latitudinal component $B_0^\theta =0$ is justified 
near the equatorial plane  and far enough from the inner magnetosphere \citep{Contopoulos_1999}.
For bodies orbiting near this plane, as circumpulsar planets probably do,}
this geometrical restriction is unimportant. 
Other models for steady state 
{axisymmetric} 
winds have been developed since then \citep{Beskin_1998,Bucciantini_2006}. 
The stationary equations of an axisymmetric perfect MHD flow}} admit a set of integrals of motion along stream lines, 
such as
the mass flux $f$ and the magnetic flux $\Psi$, which,
for cold radial flows and radial polo\"idal fields, are defined by:
\begin{eqnarray}
f&=&\gamma_0 \rho'_0 v_O^r r^2, \\ \label{eq_def_Psi}
\Psi&=&r^2 B_0^r,
\end{eqnarray}
{where all quantities refer to the unperturbed wind as seen in the  
observer's frame.}
The MHD approximation $\ve{E}_0+\ve{v}_0 \times \ve{B}_0=0$ and the Faraday equation imply \citep{Mestel_1961}: 
\begin{equation} \label{eq_approx_mhd}
B_0^\phi=B_0^r \frac{v_0^\phi -\Omega_* r}{v_0^r}.
\end{equation}
{ {In order to judge whether the planet is in superalfv\'enic motion
with respect to the wind, we should assess  whether the modulus 
of its velocity in the wind's frame, which equals that of the wind in the 
observer's frame, $v_0$, is faster or not than the propagation speed 
in the wind's frame, $V'_A$, defined by Eq. (3). The square of the ratio 
of these two velocities is given, for an asymptotically radial wind, by~:}}
\begin{equation}
{M'_A}^2 =  \left(\frac{v_{0}^r}{V'_A}\right)^2  =  \left(\frac{v_{0}^r}{c}\right)^2 
\left( 1 + \frac{c^2}{{c'_A}^2 } \right).
\end{equation}
{ {The wind being supposedly radial, $v_0^\phi$ vanishes. This simplifies equation (6) which also shows that at
distances much larger than the light cylinder radius, 
the radial component of the magnetic field can be neglected compared
to the azimuthal one. The Alfv\'en speed $c_A$ calculated in the observer's frame is given, 
using equations (4) and (5) to express $B_{0}^r$ and $\rho^0$, by:
\begin{equation}
\frac{c_A^2}{c^2}  = \frac{\Omega_*^2 \Psi^2 }{\mu_0 f c^3} \ \frac{c}{v_0^r}.
\end{equation}
The square of the Alfv\'en speed $c'_A$ in the wind's frame is  a factor $\gamma_0$ smaller than the
square of the Alfv\'en speed $c_A$ in the observer's frame
because $\rho_0 = \gamma_0 \rho'_0$ and $B_{\phi 0} = \gamma_0 B'_{\phi 0}$. The latter relation 
is a result of the vanishing of
the electric field in the wind's frame and of the velocity ${\mathbf{v}}_0$
being perpendicular to the azimuthal field. Then:}}
\begin{equation} \label{eq_mach_a}
{M'_A}^2 =  \left(\frac{v_{0}^r}{c}\right)^2 \left[1 + \frac{\gamma_0}{\sigma_0} \left(\frac{v_0^r}{c}\right)\right],
\end{equation}
{where $\sigma_0$ is the magnetization parameter:}
\begin{equation} \label{eq_def_sigma}
\sigma_0 = \frac{\Omega_*^2 \Psi^2 }{\mu_0 f c^3}.
\end{equation}
{ {In highly relativistic Poynting-flux-dominated outflows, as is expected for pulsar winds, 
$\sigma_0 \gg 1$.  The models show that asymptotically the Lorentz factor $\gamma_0$ 
approaches $\sigma_0^{1/3}$ and thus  $(v_0/c)^2$ approaches $(1 - \sigma_0^{-2/3})$. 
From equation (\ref{eq_mach_a}),
it is found that, to lowest order in an expansion in $\sigma_0^{-1}$, the asymptotic value of 
${M'_A}$ is $(1 - 1/(2\sigma_0^{4/3}))$. Therefore the pulsar wind
remains slower than the total Alfv\'en speed (3).}}

Since this result applies at distances much larger than the
light cylinder radius, it is essentially valid wherever planets may be found orbiting. We therefore consider 
that the planets detected around pulsars orbit in a sub-Alfv\'enic relativistic pulsar wind.

In the numerical computations, we  are less subtle and state that at the planetary distances, $v_0 \sim c$ and $V_A \sim c$, every time it makes sense. Then, in first approximation,
the unipolar inductor electric potential drop $U$ (along the $\theta$ axis, i.e. perpendicular to the orbital plane) is
\begin{equation} \label{eq_unipolar_pour_application_numerique}
U = 2 R_P E_0=2 R_P E_0^\theta = 2 R_P v_0^r B_0^\phi = \frac{2 R_P \Omega_* \Psi}{r}.
\end{equation}
The Fig. \ref{fig_inducteur_unipolaire} provides a representation of the geometric configuration of the unipolar inductor.

The rotation period of the pulsar PSR 1257+12 is $P=0,006$s, which corresponds to
$\Omega_*$  = 1010,397  rad.s$^{-1}$ and its surface  magnetic field is estimated 
to $B_* \simeq  8.8 \times 10^{8}  $G  \citep{Taylor_2000}. 
We assume a star radius $R_*=10$ km and a mass $M_* =1.4 M_{\odot}$ 
(see Table \ref{table_input_pulsars}). 
{{ 
Data concerning
this planet can be read in table \ref{table_input_planetes}. We assume
}}  
an Earth-like density of 5000 kg.m$^{-3}$. 
The flux $\Psi=R_*^2 B_*=8.8 \times 10^{12}$ Wb. 
The semi-major axis, 
the planetary radius $R_P= (3 M_P / 4 \pi \rho_P)^{1/3}$ and an estimate of $U$ 
are given for each planet "a", "b", "c" orbiting this pulsar in 
Tables \ref{table_input_planetes} and \ref{table_application_planetes}. 
It can be seen that the inductor electric potential drop $U$ 
(from pole to pole along the planet) is of the order of $10^{12}$ V. 

For the pulsar PSR 1620-26, the rotation period is $P=0,011$s, $\Omega_*$  = 567  rad.s$^{-1}$, the surface 
magnetic field is estimated to $B_* = 3. \times 10^{9}  $G \citep{Taylor_2000}. Still assuming that $R_*=10$ km, we find that
$\Psi =3 \times 10^{13}$ Wb. This pulsar has a white dwarf companion star. The neutron star mass 
can be estimated to $M_* =1.35 M_{\odot}$ \citep{Thorsett_1999,Sigurdsson_2003_psr_planet}. 
The planet is more distant from its star than in the case of PSR 1257+12 
but the planetary radius is larger. The resulting potential drop $U$ is still of the same order of magnitude. 

We show in sections \ref{sec_mhd}--\ref{sec_current} that a planet in the wind generates two current systems 
that propagate far in space, forming a so-called Alfv\'en wing.


\section{Equations of special-relativistic ideal MHD} 
\label{sec_mhd}

{ {In this section, the equations up to Eq. (\ref{eq_induction_0}) are general to special relativity, and valid for quantities defined in any inertial reference frame. We use notations without prime and subscript. After Eq. (\ref{eq_induction_0}), quantities without prime and subscript refer, as in the previous sections, only to the observer's frame $R_O$ defined at the begining of section  \ref{sec_unipolar_inductor}. }}

A condition for MHD to be valid
is that the typical length scale $L$ 
relevant to the flow be much larger than the particles Larmor radii $\rho_L = \gamma v_{\perp} / \omega_c$, 
where $v_{\perp}$ is the typical velocity perpendicular to the magnetic field
in the fluid's frame, $\omega_c$  the gyrofrequency and $\gamma$ the particle's Lorentz factor in the same frame. 
The energy of particles in a pulsar wind is {\it a priori}  very high since
the wind's bulk Lorentz factor $\gamma_0$ may be as large as $10^{5}$ to $10^{7}$. 
If however, in the wind's frame,  a significant part of this energy resided
in perpendicular motions, 
particles would very quickly
loose it by synchrotron radiation, even in a moderate magnetic field.
Therefore, we may consider the pulsar wind particles to have  negligible 
Larmor radii in the rest frame of the wind's bulk flow,
so that MHD is applicable on almost any scale.

The fourth component of position in space time is $x^4 = ct$, $t$ being
the time measured in seconds in the given reference frame and $c$ the speed of light. 
Greek indices  label four-dimensional
coordinates and components and latin indices label Euclidean three-dimensional ones.
The metric tensor of Minkowskian space is diagonal, with components $\eta_{44} = + 1$ and
$\eta_{11} = \eta_{22} = \eta_{33} = -1$. $\nabla_\mu$ designates
the partial derivative with respect to the space-time coordinate $x^\mu$.  
The notation ${\boldsymbol{\nabla}}$ designates the three-dimensional nabla operator. 
We use the dummy index rule.
Special-relativistic MHD equations associate fluid equations with
Maxwell's equations, in which the displacement current 
and Poisson's equation should be retained.
The fluid equations consist of a conservation equation
for particle number, valid in the absence of reactions among particle species, 
and of the four components of the energy and momentum conservation equations
The law of conservation of particle number is written as:
\begin{equation}
\nabla_\mu (n_{F} \,  u^\mu) = 0\, .
\label{numberconserv}
\end{equation}
The density $n_{F}$ is the proper spatial number density of particles, that is, their density 
measured in the instantaneous rest frame of the fluid.
The four components $u^\mu$ are those of the dimensionless four-velocity of the fluid
in the considered rest frame:
\begin{equation}
u^{\mu} = (\gamma \, {\ve{v}}/c \, , \, \gamma) \, .
\label{quadrivitesse}
\end{equation}
The Lorentz factor $\gamma$ refers here to the bulk fluid motion.
The number density in the observer's frame is the fourth component of the density-flux four-vector,
$n_{F} \gamma$.
The equations of conservation of energy and momentum of matter
are lumped in the four-tensorial equation:
\begin{equation}
\nabla_\mu T_m^{\mu \nu} = f_{em}^\nu \, .
\label{conservTmunumatt}
\end{equation}
The $T_m^{\mu \nu}$'s are the components of the energy-momentum tensor of matter and
the components $f_{em}^\nu$ are those of the electromagnetic force density four-vector. 
By using Maxwell's equations, this four-vector can be written in conservative form
and equation (\ref{conservTmunumatt}) can be given the form: 
\begin{equation}
\nabla_\mu (T_m^{\mu \nu} + T_{em}^{\mu \nu}) = 0 \, .
\label{conservTmunu}
\end{equation}
The $ T_{em}^{\mu \nu}$'s are the components of the electromagnetic
energy-momentum tensor. They can be expressed in
terms of the electromagnetic field strength tensor $F$ or in terms of the electric and magnetic fields
observed in the chosen reference frame. Its
time-time component is the electromagnetic energy density,
its space-time components are the three components of the Poynting vector
divided by $c$ and its space-space components form a second rank tensor
of the three-dimensional Euclidean space, the  Maxwell stress tensor:
\begin{equation}
{\overline{\overline{\mathbf{M}}}}= \left(\frac{\varepsilon_o E^2}{2} + \frac{B^2}{2\mu_0}\right)
\, {\overline{\overline{\boldsymbol{\delta}}}}
- \varepsilon_o {\overline{\overline{{\mathbf{E}} {\mathbf{E}} }}}
- \frac{{\overline{\overline{ {\mathbf{B}} {\mathbf{B}} }}}   }{\mu_0} \, .
\label{Maxwellstresstensor}
\end{equation}
The symbol $\boldsymbol{\delta}$ represents the second rank unit tensor.
The matter energy-momentum tensor of a cold pressureless fluid can be written, in
the absence of non-ideal effects such as viscosity, as:
\begin{equation}
T_m^{\mu \nu} = \rho_{F} c^2 u^{\mu}u^{\nu} \, .
\label{Tmunumattfroide}
\end{equation}
In the absence of internal heat, the number conservation equation (\ref{numberconserv}) reduces to a conservation
equation for proper mass, since in this case $\rho_{F} = m \, n_{F}$, 
$m$ being the rest mass of particles. Equation (\ref{numberconserv}) can then be written in
three-dimensional notations as:
\begin{equation}
\frac{\partial \, \gamma \rho_{F}}{\partial t} + {\mathrm{div}} \, (\gamma \rho_{F} {\ve{v}}) = 0 \, .
\label{conservnumbercold}
\end{equation}
Here, we need not solve the energy conservation equation because the medium is regarded as cold.
We only consider the spatial components of the equivalent equations 
(\ref{conservTmunumatt}) or (\ref{conservTmunu}).
From equation (\ref{conservTmunumatt}) we get, denoting the charge and current density by
$\rho_e$ and ${\mathbf{j}}$ respectively:
\begin{equation}
\frac{\partial}{\partial t} \left(\gamma^2\rho_{F} \, {\mathbf{v}} \right)
+ {\mathrm{div}} \, \left( \gamma^2 \rho_{F} \, {\overline{\overline{{\mathbf{v}} {\mathbf{v}} }}} \right) =
\rho_e {\mathbf{E}} + {\mathbf{j}} \times {\mathbf{B}} \, .
\label{motionavecfem}
\end{equation}
From equation (\ref{conservTmunu}) we get the equivalent equation:
\begin{equation}
\frac{\partial}{\partial t} \left(\gamma^2\rho_{F} \, {\mathbf{v}}
+ \frac{{\mathbf{E}}\times {\mathbf{B}}  }{\mu_0 c^2}\right)
+ {\mathrm{div}} \, \left( \gamma^2 \rho_{F} \, {\overline{\overline{{\mathbf{v}} {\mathbf{v}} }}} +
{\overline{\overline{\mathbf{M} }}}\right) = 0 \, .
\label{mouv3Drelativiste}
\end{equation}
The components of the Maxwell stress tensor ${\overline{\overline{\mathbf{M} }}}$ may
be transformed to account for the perfect MHD relation 
\begin{equation}
{\mathbf{E}} + {\mathbf{v}} \times {\mathbf{B}} = 0 \, . 
\label{perfectMHDrelation}
\end{equation}
The components of the electric field and those of
the Maxwell stress tensor are easily calculated in a frame where the $x$-axis is taken to be along
the direction of the fluid velocity ${\mathbf{v}}$. Some algebra then yields the following expression,
where the indices $t$ (transverse) and $l$ (longitudinal) refer to a component of a vector perpendicular or
parallel to the velocity ${\mathbf{v}}$ of the fluid:
\begin{equation}
{\overline{\overline{\mathbf{M}}}} =
\frac{B_l^2}{2 \mu_0}
\ {\overline{\overline{\boldsymbol{\delta} }}}
+ \frac{1}{\gamma^2} \frac{B_t^2}{2 \mu_0} \ {\overline{\overline{\boldsymbol{\delta} }}}
- \frac{{\overline{\overline{ {\mathbf{B}} {\mathbf{B}} }}}   }{\mu_0}
+ \frac{B_t^2}{\mu_0} \,  \frac{{\overline{\overline{ {\mathbf{v}} {\mathbf{v}}   }}} }{c^2}
+ \frac{v^2}{c^2} \, \frac{{\overline{\overline{ {\mathbf{B}}_t {\mathbf{B}}_t }}} }{\mu_0} \, .
\label{tenseurMaxwelllabo}
\end{equation}
The magnetic field evolution equation, deduced from Faraday's equation 
and the perfect MHD relation, writes:
\begin{equation} 
\frac{\partial {\ve{B}} }{\partial t} = {\mathrm{curl}} ({\ve{v}} \times {\ve{B}}) \, .
\label{eq_induction_0}
\end{equation}
The non-relativistic theory of \citet{Neubauer_1980}
considers only the Alfv\'enic wake of the satellite.
Indeed, fast MHD disturbances propagate isotropically in the low-$\beta_P$ limit
and decrease in amplitude with distance from
the source ($\beta_P$ is the ratio of the plasma pressure to the magnetic pressure). 
Such disturbances do not create any concentrated current system.
The slow mode propagates in a low-$\beta_P$ plasma much slower than Alf\'enic 
perturbations and the associated disturbances, though channelled by the magnetic field, 
soon become spatially separated from the Alf\'enic wake. Actually, slow mode
perturbations barely propagate at all in a cold pulsar wind and
carry negligible current.
The effects of compressive perturbations in the non-relativistic situation have
been discussed by \citet{Wright_1990}
who have shown that compressive plasma wave modes, though
necessarily excited, contribute one order of magnitude less to the 
current flow and to the energy budget than shear Alfv\'en perturbations do.
We therefore follow Neubauer in concentrating on purely Alfv\'enic motions. 
We assume that the fluid motions triggered when the planet passes by
are sub-relativistic in the rest frame $R_W$ of
the unperturbed wind and that the changes of the formal Alfv\'en velocity
are similarly sub-relativistic. We do not assume however that  the formal Alfv\'en speed 
calculated from the total field, $B'/(\mu_0 \rho')^{1/2}$, is 
less than the speed of light. 
Our assumption of non-relativistic Alfv\'enic motions in $R_W$ implies that
${\mathrm{div}}\, {\ve{v}}' = 0$. 
It does not follow however that the same relation also holds true in the observer's frame $R_O$ because
the wind flow is relativistic in this frame. 
In $R_W$, the electric terms of the Maxwell stress tensor are by of order $v'^2/c^2$ less
than the magnetic terms and can be neglected. 
Accounting for the relation ${\mathrm{div}}\, {\ve{v}}' = 0$, equation (\ref{eq_induction_0}) can be written as:
\begin{equation}
\frac{d {\mathbf{B}}' }{d t'} - \left({\mathbf{B}}' \cdot {\boldsymbol{\nabla}}'\right) {\mathbf{v}}' = 0
 \, .
\label{inducdsurdt}
\end{equation} 
Considering Eqs. (\ref{perfectMHDrelation}) and (\ref{tenseurMaxwelllabo}),
the equation of motion (\ref{mouv3Drelativiste}) becomes, in $R_W$,
\begin{eqnarray}
\nonumber
\frac{\mathrm{d}}{ {\mathrm{d}} t'} &&\left(\rho' {\mathbf{v}'} +
\frac{B'^2 {\mathbf{v}'}}{\mu_o c^2}  - \frac{({\mathbf{v}'}\cdot {\mathbf{B}'}) \, {\mathbf{B}'}}{\mu_o c^2} \right)
\\ 
&&- \Big({\mathbf{B}'} \cdot {\boldsymbol{\nabla}'} \Big) \left( \frac{{\mathbf{B}'}}{\mu_o } 
+  \frac{({\mathbf{v}'}\cdot {\mathbf{B}'})\, {\mathbf{v}'} }{\mu_o c^2}\!\right)
= 0. 
\label{motionAlfrefvent}
\end{eqnarray}


\section{An Alfv\'enic first integral} \label{sec_invariant}
The computation of the non relativistic Alfv\'en wing given by \citet{Neubauer_1980} 
is based upon the fact that, {{in simple non-linear Alfv\'enic wave motions, the velocity}} 
$\ve{V}_s = \ve{v} - s \ \ve{c}_A$
is a first integral. In this relation, 
${\mathbf{v}}$ is the fluid's velocity,
$\ve{c}_A$ the vectorial Alfv\'en velocity associated with the perturbed field, and the sign $s=\pm 1$
{{depends on the sense
of propagation of the perturbation}}. 
This relation can be transposed in a differential form as $d \ve{B}= k d\ve{v}$, where $k = (\mu_0 \rho)^{1/2}$.
Moreover, the modulus of the magnetic field is a time invariant and, because 
of the uniform boundary conditions $B=B_0$, this modulus is constant over the whole space.

As Neubauer, we look for a solution where $d \ve{B}'= k d\ve{v}'$ and where $B'$ is constant.
Then, setting $\ve{b}'=\ve{B}'/B'$ and 
\begin{equation} \label{lambda_dns_RW}
\lambda = (\mu_0 \rho'+B'^2/c^2)^{1/2},
\end{equation}
we can write the Eqs. (\ref{inducdsurdt}) and (\ref{motionAlfrefvent}) 
in the form
\begin{eqnarray}
&& \lambda^2 \frac{{d} \ve{v}'}{ {{d}} t'} \!-B'^2 (\ve{b}'\cdot\nabla') \ve{b}' \!
-\!\frac{B'^2}{c^2} \left[\frac{d v'_\parallel \ve{b}'}{d t'} \! + \! (\ve{b}'\cdot\nabla') (v'_\parallel \ve{v}')\right]\!=0 
\label{motionAlfrefvent_2}
\\ 
\label{inducdsurdt_2}
&& \frac{d {\mathbf{b}}' }{d t'} - ({\mathbf{b}}' \cdot {\boldsymbol{\nabla}}') {\mathbf{v}}' = 0,
\end{eqnarray}
where $v'_\parallel = \ve{v}' \cdot \ve{b}'$.
The two first terms of Eq. (\ref{motionAlfrefvent_2}) are analogous to those of classical MHD, when 
$V_A << c$. The two others are specific to fast variations of the electric field, when the displacement current is taken into account.
We solve this system in the following way: 
we first ignore the last two terms of Eq. (\ref{motionAlfrefvent_2}) and solve the resulting system
(\ref{motionAlfrefvent_2})--(\ref{inducdsurdt_2}), which can then be written as:
\begin{eqnarray} \label{motionAlfrefvent_reduced}
&& \lambda^2 \frac{\mathrm{d} \ve{v}'}{ {\mathrm{d}} t'} -B'^2 (\ve{b}'\cdot\nabla') \ve{b}'=0, 
\\ 
\label{inducdsurdt_3}
&& \frac{d {\mathbf{b}}' }{d t'} - ({\mathbf{b}}' \cdot {\boldsymbol{\nabla}}') {\mathbf{v}}' = 0.
\end{eqnarray}
We then check whether the last two terms in Eq. (\ref{motionAlfrefvent_2}) vanish or are negligible.
These terms are:
\begin{equation} \label{motionAlfrefvent_complement} 
\frac{B'^2}{c^2} \left[ \frac{d v'_\parallel \ve{b}'}{d t'} +  (\ve{b}'\cdot\nabla') (v'_\parallel \ve{v}')\right] \, .			
\end{equation}
Considering Eqs. (\ref{motionAlfrefvent_reduced}) and (\ref{inducdsurdt_3}), 
the proportionality of $d \ve{B}'$ and $d\ve{v}'$ is obtained for $k=s\lambda$, and these two equations become equivalent to 
\begin{eqnarray} \label{solution_reduite_1} 
&&B' d\ve{b}' = s \lambda d\ve{v}' \\
\label{solution_reduite_2} 
&&s \lambda \frac{d {\mathbf{b}}' }{d t'} - B' (\ve{b}'\cdot\nabla') \ve{b}'=0.
\end{eqnarray}
Eq. (\ref{solution_reduite_1}) has a first integral 
$\ve{V}'_s$,
\begin{equation} \label{invardansRW}
\ve{V}'_s = \ve{v}' - s  \, \frac{\ve{B}'}{(\mu_0 \rho'+B'^2/c^2)^{1/2}}.
\end{equation}
In the region not perturbed by the planet, $\ve{v}'=0$, and $\ve{B}'=\ve{B}'_0$. Therefore, 
\begin{eqnarray} \label{valeur_invariant_prime}
&&\ve{V}'_s = -s \frac{\ve{B}'_0}{\lambda},\\
&&\ve{v}'=  \frac{s}{\lambda}(\ve{B}'-\ve{B}'_0),\\ \label{solution_simple_4}
&&v'_\parallel = \frac{s}{\lambda} (B' - \ve{b}' \cdot \ve{B}'_0).
\end{eqnarray}
In order of magnitude, $\lambda \sim B'/c$, and the approximation $v'<<c$ implies that 
$\|\ve{B}'-\ve{B}'_0\|<<B'_0$. Therefore, this is equivalent to an hypothesis of linear perturbation.
This remark also holds for the computations of \cite{Neubauer_1980}.
At this stage, it is necessary to evaluate the terms in Eq. (\ref{motionAlfrefvent_complement}).
Using the above relations, these terms are 
\begin{equation}
-s \, \frac{B'^2}{c} \left( \frac{d \ve{b}'}{d t'} \cdot \frac{\ve{v}'}{c}\right) \ve{b}'_0 \, , 
\label{equa36}
\end{equation}
where $\ve{b}'_0$ is the direction of the unperturbed magnetic field in the wind's frame of reference.
The quantity in Eq. (\ref{equa36}) does not vanish but it is negligible, as long as $v'<<c$, 
in comparison to the terms in Eq. (\ref{motionAlfrefvent_reduced}), which are of the order of $(B'^2/c) d \ve{b}'/dt'$.
Therefore, in the linear approximation, the solution 
given by Eqs. (\ref{solution_reduite_1}-\ref{solution_simple_4}) is correct.

The first integral of Eq. (\ref{invardansRW}) must now be computed in the observer's frame 
of reference $R_O$. It will be used in the next section to derive the value of the electric current 
associated to the Alfv\'en wings.
{ { 
Let us write the velocity and the magnetic field as the sum of a longitudinal component, 
parallel to $\ve{v}_0$, and a transverse vector component.
\begin{eqnarray}
{\ve{B}}_l &=& \frac{ {\overline{\overline{{\ve{v}}_0 {\ve{v}}_0}}} }{v_0^2 } \cdot {\ve{B}},\\
{\ve{B}}_t &=& 
 \, \left( {\overline{\overline{ \boldsymbol{\delta} }}}
-  \frac{ {\overline{\overline{ {\ve{v}}_0 {\ve{v}}_0 }}} }{v_0^2 }\right)
\cdot {\ve{B}}.
\end{eqnarray}
The transforms of the velocity and of the magnetic field are 
\begin{eqnarray}
{\mathbf{v}}' &=& \frac{
{\mathbf{v}}_{l} - {\mathbf{v}}_0 + {\mathbf{v}}_{t}/\gamma_0
}
{
1 - {{\mathbf{v}} \cdot {\mathbf{v}}_0}/{c^2}}, 
\label{transfovitesse}
\\
 {\mathbf{B}}_{l}' &=&   {\mathbf{B}}_{l},
\label{changeBparallel} \\
{\mathbf{B}}_{t}'  &=& \gamma_0 \left( {\mathbf{B}}_{t} \left( 1 - \frac{{\mathbf{v}} \cdot {\mathbf{v}}_0}{c^2}\right)
+ \frac{{\mathbf{B}} \cdot {\mathbf{v}}_0}{c^2} {\mathbf{v}}_{t} \right),
\label{changeBperp}
\end{eqnarray}
where $\gamma_0$ is the Lorentz factor of the unperturbed wind.
For the transform of the magnetic field, Eq. (\ref{perfectMHDrelation}) has been taken into account.
The Eq. (\ref{invardansRW}) becomes
\begin{eqnarray} \nonumber
{\mathbf{V}}'_s &=& \frac{ 
{\mathbf{v}}_l - {\mathbf{v}}_0 + {\mathbf{v}}_t/\gamma_0 
}{
1 - \frac{{\mathbf{v}} \cdot {\mathbf{v}}_0}{c^2}   
}
\\
& &- \, \frac{s}{\lambda} 
\, \left( \gamma_0 \left( {\mathbf{B}}_{t} \left(1 - \frac{{\mathbf{v}} \cdot {\mathbf{v}}_0}{c^2} \right) 
+ \frac{{\mathbf{B}} \cdot {\mathbf{v}}_0}{c^2} \, {\mathbf{v}}_t\right)  + {\mathbf{B}}_{l} \right).
\label{invarRObrut}
\end{eqnarray}
For a first order development, we note ${\mathbf{v}_1}$ the velocity perturbation and ${\mathbf{B}_1}$ the magnetic perturbation,
\begin{equation}
{\mathbf{v}} = {\mathbf{v}}_0 + {\mathbf{v}_1} \qquad \qquad {\mathbf{B}} = {\mathbf{B}}_0 + {\mathbf{B}_1}.
\label{linearRO}
\end{equation}  
To the first order, Eq. (\ref{invarRObrut}) becomes
\begin{eqnarray}
{\mathbf{V}}'_s \ &=&  \gamma_0^2 \left({\mathbf{v}}_{1 l} + \frac{{\mathbf{v}}_{1 t}}{\gamma_0}  \right)
- \, \frac{s}{\lambda} \left({\mathbf{B}}_{0 l} + {\mathbf{B}}_{1 l} \right)
\nonumber \\
&-& \ \frac{s \gamma_0 }{\lambda} \ \left(
\frac{1}{\gamma_0^2} \left( {\mathbf{B}}_{0 t} + {\mathbf{B}}_{1 t} \right)  
- {\mathbf{B}}_{0 t} \frac{{\mathbf{v}}_0 \cdot {\mathbf{v}_1}}{c^2} +
\frac{{\mathbf{B}}_0 \cdot {\mathbf{v}}_0 }{c^2} \, {\mathbf{v}}_{1 t} \right).
\label{invarROlinearise}
\end{eqnarray}
which can be rewritten
\[ 
\frac{{\mathbf{V}}'_s}{\gamma_0^2} = 
{\mathbf{v}}_{1 l} + \left(1 - \frac{s}{\lambda} \, \frac{{\mathbf{B}}\cdot {\mathbf{v}}_0 }{c^2} \right) 
\, \frac{{\mathbf{v}}_{1 t}}{\gamma_0} - \frac{s}{\lambda \gamma_0^2} \, {\mathbf{B}}_{l}
- \frac{s}{\lambda \gamma_0} \,\left( \frac{1}{\gamma_0^2} - \frac{{\mathbf{v}_1}\cdot {\mathbf{v}}_0 }{c^2}\right) 
{\mathbf{B}}_{t}.
\] 
Considering the purely geometric relation 
\begin{equation}
\frac{s}{\lambda} \ \frac{{\mathbf{B}}\cdot {\mathbf{v}}_0 }{c^2} \ {\mathbf{v}}_{1 l} =
\frac{s}{\lambda} \ \frac{{\mathbf{v}_1} \cdot {\mathbf{v}}_0 }{c^2} \  {\mathbf{B}}_{l},
\end{equation}
we have
\begin{eqnarray}
\frac{{\mathbf{V}}'_s}{\gamma_0^2} = && +
\left(1 - \frac{s}{\lambda} \, \frac{{\mathbf{B}}\cdot {\mathbf{v}}_0 }{c^2} \right) \ {\mathbf{v}}_{1 l}
\ - \, \frac{s}{\lambda} \, \left(1 - \frac{v_0^2}{c^2} - \frac{{\mathbf{v}}_0 \cdot {\mathbf{v}_1}}{c^2}\right) 
\ {\mathbf{B}}_{l}
\nonumber \\
&& + \left(1 - \frac{s}{\lambda} \, \frac{{\mathbf{B}}\cdot {\mathbf{v}}_0 }{c^2} \right) 
\frac{{\mathbf{v}}_{1 t}}{\gamma_0}
- \frac{s}{\lambda} \, \left(1 - \frac{v_0^2}{c^2} - \frac{{\mathbf{v}}_0 \cdot {\mathbf{v}_1}}{c^2}\right) 
\frac{{\mathbf{B}}_{t}}{\gamma_0}.
\label{invarROavanthomothetie}
\end{eqnarray}
As this vector is constant, its longitudinal and transverse parts relatively to the constant vector ${\mathbf{v}}_0$ are also two distinct constant entities. We can recombine the longitudinal and transverse components in the following way,
\[ 
\left(1 - \frac{s}{\lambda} \, \frac{{\mathbf{B}}\cdot {\mathbf{v}}_0 }{c^2} \right) {\mathbf{v}_1}
- \frac{s}{\lambda} \, \left(1 - \frac{v_0^2}{c^2} - \frac{{\mathbf{v}}_0 \cdot {\mathbf{v}_1}}{c^2}\right) 
{\mathbf{B}} =
\left( \frac{{\mathbf{V}}'_s}{\gamma_0^2}\right)_l + \gamma_0 \left( \frac{{\mathbf{V}}'_s}{\gamma_0^2}\right)_t.
\label{invarROapreshomothetie}
\] 
Then we can define a new first integral vector,
\begin{equation}
{\mathbf{v}} \, - \, \frac{s}{\alpha} \  {
{\mathbf{B}} \ \left(1 - \frac{v_0^2}{c^2} - \frac{{\mathbf{v}}_0 \cdot {\mathbf{v}_1}}{c^2}\right)
}
=  \ {\mathbf{v}}_0 \ + \ \frac{
{\mathbf{V}}'_{sl}/\gamma_0^2 + {\mathbf{V}}'_{st}/\gamma_0
}{
1 - \frac{s}{\lambda} \, \frac{{\mathbf{B}}_0 \cdot {\mathbf{v}}_0 }{c^2} 
},
\label{invardansR0}
\end{equation}
where the constant number $\alpha$ is defined by
\begin{equation}
\alpha = \lambda  -s \frac{{\mathbf{B}}_0 \cdot {\mathbf{v}}_0 }{c^2}.
\end{equation}
In spite of a greater complexity, this vector presents some analogy with the first integral found in the non relativistic case by Neubauer (written in the first lines of the present section). This 
will be used in the next section to derive the current flowing along the Alfv\'en wing, in a similar way to those developed in \cite{Neubauer_1980}.
The parameter $\lambda$, given in Eq. (\ref{lambda_dns_RW}), can be expressed as a function of the unperturbed wind parameters,
\begin{equation}
\lambda = \left[\mu_0 \rho_0'+c^{-2} {B}_{l 0}^2+ c^{-2}  \gamma_0^{-2} B_{t 0}^2 \right] ^{1/2}.
\end{equation}
Since $\nabla \cdot \ve{v}'=0$, the density $\rho'$ is invariant along any line of flow;
that is why we have noted it $\rho_0'$ instead.
We set 
\begin{eqnarray}
c_A^2&=&B_0^2/\mu_0  \rho_0' ,
\\ c_{A l}^2&=&{B}_{l 0}^2/\mu_0  \rho_0', \\
 c_{A t}^2&=&B_{t 0}^2/\mu_0  \rho_0'.
\end{eqnarray}
These are mixtures of the magnetic field in the frame $R_O$ of the observer 
and of the density in the proper frame $R_W$ of the wind in terms of which the factor $\lambda$  can be written as:
\begin{equation}
\lambda = \sqrt{\mu_0  \rho_0'} \ \left(1+ \frac{c_{A l}^2  \gamma_0^2+c_{A t}^2}{ \gamma_0^2 c^2}\right) ^{1/2}.
\end{equation}
}} 
\section{Current carried by an Alfv\'en wing} \label{sec_current}
\citet{Neubauer_1980} has shown that the electric current density carried by an Alfv\'en wing 
is related to the divergence of the electric field through the relation 
$\mu_0 \, \ve{j} \cdot \ve V_s= \Sigma \, {\mathrm{div}}\, {\mathbf{E}}$
where the conductance $\Sigma = (\mu_0 c_A)^{-1} (1+ M_A^2 -2 s M_A \sin \theta)^{-1/2}$, $c_A$ being the 
classical Alfv\'en velocity, and $M_A=v_0/c_A$. In this section, 
we derive an analogous relation for the relativistic plasma flow, {{keeping our previous assumptions.
From the perfect MHD relation (\ref{perfectMHDrelation}) we calculate ${\mathrm{div}} \, {\mathbf{E}}$ in
the observer's frame $R_O$:
\begin{equation}
{\mathrm{div}} \, {\mathbf{E}} = {\mathbf{v}} \cdot \, {\mathrm{curl}} \, {\mathbf{B}} 
- {\mathbf{B}} \cdot \, {\mathrm{curl}} \, {\mathbf{v}}.
\end{equation}
{ {
To the first order,
\begin{equation} \label{divE_depart_lin}
{\mathrm{div}} \, {\mathbf{E}} = {\mathbf{v}_0} \cdot \, {\mathrm{curl}} \, {\mathbf{B}_1} 
- {\mathbf{B}_0} \cdot \, {\mathrm{curl}} \, {\mathbf{v}_1}.
\end{equation}
From Eq. (\ref{invardansR0}), the curl of the velocity is 
\begin{eqnarray} \label{curl_v}
{\mathrm{curl}} \, {\mathbf{v}_1} &=& \frac{s}{\gamma_0^2 \alpha} \, {\mathrm{curl}} \, {\mathbf{B}_1}
-\frac{s}{\alpha c^2} {\mathrm{curl}} \, [{\mathbf{B}_0} ({\mathbf{v}_0} \cdot {\mathbf{v}_1})]
\\
&=& \frac{s}{\alpha \gamma_0^2} \, {\mathrm{curl}} \, {\mathbf{B}_1} -\frac{s}{\alpha c^2} {\mathbf{N}} \times {\mathbf{B}_0},
\end{eqnarray}
where 
\begin{equation}
{\mathbf{N}}= ({\mathbf{v}_0} \cdot \nabla) {\mathbf{v}_1} + {\mathbf{v}_0} \times {\mathrm{curl}} \, {\mathbf{v}_1}.
\end{equation}
Including Eq. (\ref{curl_v}) in Eq. (\ref{divE_depart_lin}), we find
\begin{equation}
{\mathrm{div}} {\mathbf{E}}= \left({\mathbf{v}_0}  - \frac{s \, {\mathbf{B}_0} }{ \gamma_0^2 \alpha} \right)
\cdot {\mathrm{curl}}\, {\mathbf{B}}.
\label{A} 
\end{equation}
For further convenience, we note 
\begin{equation}
 {\mathbf{U}_s}= {\mathbf{v}_0}  - \frac{s \, {\mathbf{B}_0} }{ \gamma_0^2 \alpha}.
\label{def_Us} 
\end{equation}
In $R_O$, the Alfv\'en wave is stationary, therefore the partial time derivatives are null, and Ampere's equation is simply:
\begin{equation}
{\mathrm{curl}}\, {\mathbf{B}}
= \mu_0 {\mathbf{j}}.
\end{equation}
Then, from Eq.(\ref{A}), 
\begin{equation}
{\mathrm{div}} {\mathbf{E}}= \mu_0 \, {\mathbf{j}} \cdot {\mathbf{U}}_s.
\label{AA}
\end{equation}
Let $J_s$ be the projection of the current density along the direction of the constant vector ${\mathbf{U}}_s$
of equation (\ref{invardansR0}). Equation (\ref{AA}) can be written as:
\begin{equation}
{\mathrm{div}} {\mathbf{E}}= \mu_0 \, J_s \, \vert{\mathbf{U}}_s\vert.
\end{equation}
Let $\theta = \pi/2 - ({\mathbf{B}}_0, {\mathbf{v}}_0)$ be the complement of the angle made in $R_O$ 
by the ambient magnetic field and the flow velocity. The modulus of ${\mathbf{U}}_s$ (equation (\ref{invardansR0}))
is
\begin{equation}
U_s^2 = v_0^2 - \frac{2 s}{\alpha \gamma_0^2} \, v_0B_0 \sin \theta + \frac{B_0^2}{\alpha^2 \gamma_0^4},
\end{equation}
and then:
\begin{eqnarray}
J_s &=& \Sigma_A \  {\mathrm{div}}\,  {\mathbf{E}} \label{JsSigmaA},\\
\Sigma_A &=&  \frac{   \gamma_0^2 \left[ \left( 1+\frac{c_{A l}^2  \gamma_0^2+c_{A t}^2}{ \gamma_0^2 c^2}\right)^{1/2} -s \frac{v_0 c_A}{c^2} \sin \theta \right] }{\mu_0 c_A \left( 1 + X_A^2 -2 s X_A \sin \theta \right)^{1/2} },
\label{SigmaA}\\
X_A &=& \frac{v_0 \alpha  \gamma_0^2}{B_0}= \frac{v_0}{c_A}   \gamma_0^2 \left( 1+\frac{c_{A l}^2  \gamma_0^2+c_{A t}^2}{ \gamma_0^2 c^2}\right)^{1/2} -s \frac{v_0^2 \gamma_0^2}{c^2} \sin \theta.
\end{eqnarray} 
When $c_A << c$, we are in the conditions studied by Neubauer, and we find the same result as in his paper (given in the beginning of this section). 
In the case  $\gamma_0 >> 1$ and $c_A >>c$, that is relevant for a pulsar's wind,  $X_A >>1$ and 
\begin{equation}
\Sigma_A \sim \frac{1}{\mu_0 c},
\end{equation}
}}
{{a conductance which
is associated with the impedance of vacuum, equal to
${\mathcal{R}}_\infty \equiv \mu_0 c =$ 377 Ohm.}}
The two directions of the currents flows correspond to the invariant vectors $\ve{U}_+$ and $\ve{U}_-$.The geometrical configuration of this solution is shown in Fig. \ref{fig_ailealfperp}.

Now, we can adapt quite directly the conclusions of \citet{Neubauer_1980} to the relativistic inductor. 
{{This author considers a specific model of the wake-aligned currents in the Alfv\'en wing, which
he assumes to be flowing on the surface of an infinite cylinder tangent to the planet's surface, with its axis parallel to
Alfv\'enic characteristics. Equation (\ref{JsSigmaA}) implies that for such currents
the divergence of the electric field vanishes except
on the cylinder's surface. The electric potential can then be found by solving Laplace's 
equation, assuming the electric field inside the cylinder to be constant, of intensity $E_i$.
This simple assumption is motivated by the difficulty to solve for the electromagnetic and 
flow structure in the immediate vicinity of the solid body. The free parameter}} 
$E_i$ is the electric field along
the planet caused by its ionosphere or surface internal resistance.
{{At large distances from the wake, the electric field converges to
the convection field ${\mathbf{E}}_0$ 
given by the perfect MHD relation (\ref{perfectMHDrelation})
in the unperturbed wind.
Once the electric potential is found, the magnetic field 
and current distribution in the wake can be deduced, using in particular equation (\ref{JsSigmaA}).}}  

Neubauer gives useful expressions for the total current $I$ flowing along an Alfv\'en wing 
{{and for the Joule dissipation power in the solid body, $\dot E_{J}$.}}  Writing $R_P$ for the planet's radius, he gets:
\begin{eqnarray} 
\label{eq_total_current}
I &=& 4 \, (E_0 - E_i) \, R_P \, \Sigma_A = 4 \, \left(\, \frac{\Omega_* \Psi}{r} - E_i \, \right)\,  R_P\, \Sigma_A,
\\ \label{eqdissJoule}
\dot E_{J} &=& 4 \pi \, R_P^2 \ E_i \, (E_0 - E_i) \ \Sigma_A.
\end{eqnarray}
The Joule dissipation is maximum when $E_i = E_0/2$. {{In our estimations, we shall 
use Neubauer's values for $I$ and $\dot E_{J}$. Up to unimportant numerical factors, the results 
(\ref{eq_total_current})--(\ref{eqdissJoule})
are simply obtained by considering the current $I$ to be driven in a resistive load of resistance $\Sigma_A^{-1}$
by a generator of electromotive force $U_0 = 2 E_0 R_P$ applied on two opposite sides of a planet
of internal resistance ${\mathcal{R}}_P$. Neubauer's parameter $E_i$ is related to ${\mathcal{R}}_P$ by:
\begin{equation} 
\frac{E_i}{E_0} = \frac{{\mathcal{R}}_P}{\Sigma_A^{-1} + {\mathcal{R}}_P}.
\end{equation}
It is very difficult to know what the value of ${\mathcal{R}}_P$ really is. 
We therefore regard it, or $E_i$, as unspecified parameters.  The planet's electrical resistance 
depends on its constitution, on the path of electric currents in it and on
the existence or absence of some form of ionosphere.
The conductivity of terrestrial silicate rock
is of order $\sigma_{rock} = 10^{-3}$ Mho m$^{-1}$ \citep{Cook_1973}. If it is assumed that 
the wake current closes through a layer of thickness $h$
at the surface of the planet, the electrical resistance ${\mathcal{R}}_P$ of
the latter would be of order $\sigma_{rock}^{-1} / h$, which numerically 
amounts to ${\mathcal{R}}_P \sim (h_{\mathrm{km}})^{-1}$. As soon as $h$ would be larger
than a few meters, which for this low conductivity is reached in a very short time, 
${\mathcal{R}}_P$ would be comparable to or
less than  ${\mathcal{R}}_\infty$, eventhough the rock intrinsically is a poor conductor. 
A second generation planet, which could be partly metallic,
or a partly molten body, could have lower resistances.}}




\begin{table*}
\caption{Input data about the pulsars. Rotation period and pulsation, surface magnetic field, mass (in solar masses), radius, and Goldreich-Julian current.  Source : \citep{Taylor_2000} catalog, in the SIMBAD database. For PSR 1257+12, the mass and the radius are infered from general ideas about pulsars. For PSR 1620-26, the mass has been measured, because the pulsar is in a binary system (with a white dwarf companion). } 
\label{table_input_pulsars} 
\centering 
\begin{tabular}{l c c c c c c} 
\hline\hline 
Name & $P$ (s) & $\Omega_*$ ($s^{-1}$) & $B_*$ (Gauss) & $M_*$ ($M_{\odot}$)& $R_*$ (km) & $I_{GJ}$ (A)\\ 
\hline 
PSR 1257+12 & 0.006 & 1010. & $8.8 \times 10^{8}$ & 1.4 & 10. & 4.9 $ \times 10^{12}$ \\
\hline 
PSR 1620-26 &  0.011  & 567  & $3. \times 10^{9}$  & 1.35 & 10.  & 5.4 $ \times 10^{12}$ \\
\hline 
PSR 10 ms &  0.010  &  628. & $ 10^{8}$  & 1.5 & 10. & 2.2 $ \times 10^{11}$  \\ 
\hline 
PSR 1 s &  1.  & 6.283  & $ 10^{12}$  & 1.5 &  10. & 2.2 $ \times 10^{11}$ \\
\hline 
\end{tabular}
\end{table*}

\begin{table*}
\caption{Input data concerning the planets : mass, radius, orbital period, semi-major axis, excentricity. Source : $M_{\oplus}$, $P_{orb}$, $a$, $e$ from "The Extrasolar Planets Encyclopaedia" (http://exoplanet.eu/index.php). The estimate of $R_P$ is based on the hypothesis of an average mass density $\rho_P=$ 5000 kg.m$^{-3}$.} 
\label{table_input_planetes} 
\centering 
\begin{tabular}{l c c c c c} 
\hline\hline 
Name & $M_P$ ($M_{\oplus}$) & $R_P$ ($R_{\oplus}$) & $P_{orb}$ (day) & $a$ (AU) & $e$\\ 
\hline 
PSR 1257+12 a & 0.02 & 0.28 & 25. & 0.19 & 0\\
PSR 1257+12 b & 4.3  & 1.68 & 66. & 0.36 & 0.0186\\
PSR 1257+12 c & 3.9  & 1.62 & 98. & 0.46 & 0.0252\\
\hline 
PSR 1620-26 a &  794  & 9,5  & 36367.  & 23.    \\
\hline 
PSR 1s b 10,000km & 3.5 & 1.57 & 30.& 0.21 & 0. \\
\hline 
PSR 10ms b 100 km & $2. \times 10^{-6}$   & 0.016  & 20.  &    0.16  & 0.3 \\
PSR 10ms b 1 km   & $2. \times 10^{-12}$   & $1.6 \times 10^{-4}$ & 20.   &  0.16 & 0.3  \\
\hline 
PSR 1 s b 100 km & $2. \times 10^{-6}$    & 0.016  & 20.  &    0.16 & 0.3  \\
PSR 1 s b 1 km   & $2. \times 10^{-12}$    & $1.6 \times 10^{-4}$  & 20.  &    0.16 & 0.3  \\
\hline 
\end{tabular}
\end{table*}

\begin{table*}
\caption{Electric potential drop, total electric current associated to the Alfv\'en wing. 
Electrical energy $\dot E_{J max}$ dissipated in the Alfv\'en wing.}  
\label{table_application_planetes} 
\centering 
\begin{tabular}{l r r r   } 
\hline\hline 
Name & $U$ (V)& $I_{AW}$ (A)& $\dot E_{J max}$ (W)  \\ 
\hline 
PSR 1257+12 a &  1.1 $\times 10^{12}$ & 3.0 $\times 10^{9}$ & 2.5$\times 10^{21}$    \\
PSR 1257+12 b &  3,5 $\times 10^{12}$ & 9.4 $\times 10^{9}$ & 2.5 $\times 10^{22}$   \\
PSR 1257+12 c &  2,6 $\times 10^{12}$ & 7.0 $\times 10^{9}$ & 1.4 $\times 10^{22}$   \\
\hline 
PSR 1620-26 a &  6,0 $\times 10^{11}$ & 1.5 $\times 10^{9}$ & 7 $\times 10^{20}$    \\
\hline 
PSR 1s b 10,000 km &  3.8 $\times 10^{13}$ &1.0 $10^{11}$ & 2. $\times 10^{24}$ \\
\hline
PSR 10ms b 100 km &   2.4$\times 10^{9}$ &  6.$\times 10^{6}$ & 1.2 $\times 10^{16}$   \\
PSR 10ms b 1 km   &   2.4$\times 10^{7}$ &  6.$\times 10^{4}$ & 1.2 $\times 10^{12}$   \\
\hline 
PSR 1 s b 100 km &  2.4  $\times 10^{11}$ &  6$\times 10^{8}$ & 1.2 $\times 10^{20}$   \\
PSR 1 s b 1 km   &  2.4 $\times 10^{9}$ &  6.$\times 10^{6}$ & 1.2 $\times 10^{16}$   \\
\hline 
\end{tabular}
\end{table*}

\section{Discussion and conclusion}
{ {This work shows that a planet orbiting a pulsar develops a system of Alfv\'en wings, caused by its interaction 
with the sub-Alfv\'enic Poynting-flux-dominated pulsar wind. A system of strong electric currents is set.
Although this current cannot reach the inner pulsar's magnetosphere, 
it is nevertheless interesting to compare it to the current at the origin of the 
pulsar's magnetospheric activity, the Goldreich-Julian current $J_{GJ}$. This latter current results from the 
electromotive field generated by the fast 
rotation of the highly magnetized neutron star and its surrounding magnetosphere.
For a dipole magnetic field, the Goldreich-Julian current density is 
\begin{equation}
J_{GJ} = c \epsilon_0 \Omega_* B_*,
\end{equation} 
and the total current is of the order of 
\begin{equation}
I_{GJ} \sim \pi R_{PC}^2 J_{GJ},
\end{equation} 
where $R_{PC} \sim R_*^{3/2} (\Omega_*/c)^{1/2}$ is the polar cap radius \citep{Kirk_2009}. 
Various values of the  Goldreich-Julian current $I_{GJ}$ are given in Table \ref{table_input_pulsars}.
A look at table \ref{table_application_planetes} allows for comparisons between  $I_{AW}$ and $I_{GJ}$.
It can be seen that in the case of the four known planets, the electric current in the 
Alfvén wings is smaller than the Goldreich-Julian current by three orders of magnitude. 
It is also much smaller in the case of small bodies. But a planet 
orbiting a "standard" pulsar with a typical 1 second period and $10^{12}$ G magnetic field would have an Alfvén wing current
$I_{AW}$  of similar amplitude as the Goldreich-Julian current $I_{GJ}$. 
This is not negligible, when we see that $I_{GJ}$ is the basic engine of the pulsar's electrodynamics.

The practical consequences of such a current are expected to be of two kinds : it is shown in a companion paper \citep{Mottez_2011_AWO} that it can exert an ortho-radial force upon the planet that can, if the magnetic to mechanical energy coupling is efficient enough, 
have an incidence on the orbit of small circum-pulsar objects such as asteroids or comets. The second incidence, more relevant 
to massive objects such as planets, is a possibly associated electromagnetic signature, which might be detectable.  }}

{ {Does the Alfvén wing dominates, in terms of energy, the direct mechanical action exerted by the wind on the companion~? 
To answer this question, we can compare the flux of magnetic energy, $F_M$ with the flux of mechanical energy,
$F_K$, received the companion by direct impingement. The ratio of the Poynting flux to the mechanical energy flux is  
\begin{equation}
\frac{F_M}{F_K}  = \frac{c\ (B_{0\phi}^2 /\mu_0)  }{c\ ( \gamma_0 \rho_0 c^2)}.
\end{equation}
Introducing $\sigma_0$ and $\Psi$ with the Eqs. (\ref{eq_def_Psi},\ref{eq_approx_mhd},\ref{eq_def_sigma}), and $v_{0r} \sim c$,
\begin{equation}
\frac{F_M}{F_K}   = \frac{\sigma_0}{\gamma_0}. 
\end{equation}
In the case discussed above where $\gamma_0 \sim \sigma_0^{1/3}$, and $\sigma_0 >> 1$,
\begin{equation}
\frac{F_M}{F_K}   \sim \sigma_0^{2/3} >> 1.
\end{equation}
Therefore, we expect that for a Poynting-flux-dominated wind, most of the energy exchange with the pulsar's companion comes from the magnetic field. 
The Eq. (\ref{eqdissJoule}) provides a more precise insight of what is effectively involved into the wind-companion interaction. 
Considering  $E_i = E_0$ and 
$E_0 \sim c B_\phi$,
\begin{equation}
\frac{{\dot{E}}_J }{\pi R_P^2} \approx E_0^2 \, \Sigma_A \approx \frac{c^2 B_\phi^2 }{\mu_0 c}   =
c \ \frac{B_\phi^2 }{\mu_0}.
\end{equation}
This amounts to the totality of the Poynting flux intercepted by the pulsar's companion. 
It is therefore  larger (by a factor $\sigma_0^{2/3}$) than the mechanical energy captured by direct impigement.
}}

{ {This paper rests on the fact that the wind velocity $v_0$ is slower than the Alfvén wave velocity $V_A$. Otherwise, there would  be no Alfvén wing. 
In the case of an ideal MHD radial wind, the Lorentz factor asymptotically approaches 
$\sigma_0^{1/3}$ and Eq. (\ref{eq_mach_a}) shows that the wind remains sub-Alfvénic at any distance. If however
it is formally considered that the asymptotic Lorentz factor 
$\gamma_{0 \infty}$ scales as  $\sigma^a$ instead, with an exponent $a \neq 1/3$, then
if $a>1/3$, $M_{A \infty} =1+\sigma^{a-1}-\sigma^{-2a} \sim 1+\sigma^{a-1} >1$. 
Therefore, in that case, a transition form a sub-Alfvénic to a super-Alfvénic wind occurs at a finite distance. Observations 
of the equatorial sectors of winds driving pulsar wind nebulae show lower values of the 
asymptotic magnetization \citep{Kennel_1984a,Kennel_1984b,Gaensler_2002} and various authors 
suggest that the asymptotic value of the Lorentz factor is rather $\gamma_{\infty} \sim \sigma_0$. 
\citet{Arons_2004} argues that dissipation must occur  in the asymptotic wind zone, in order to understand 
the observed high Lorentz factors and the low magnetization. \cite{Begelman_1994} show that 
when the flux tubes diverge faster than radially, the fast magnetosonic point can occur 
closer to the light cylinder, implying an even closer Alfvénic point. In such a circumstance, the existence of Alfvén wings would depend on the distance from the star to the planet, combined with the effect of a non-radial diverging wind flow, or dissipation.  

Let us come back to the hypothesis of a planet in a sub-Alfvénic wind.}}
The present work provides only orders of magnitude estimates for the emitted current. 
The behaviour of the Alfv\'en wing at close vicinity of the planet would need a more detailed study. 
We have considered here, as in many other papers concerning pulsars, 
the case of a neutron star magnetic field that is aligned with the rotation axis. 
The study of oblique rotators raises more complicated problems. 
The pulsar wind may have different properties, and carry, even in the equatorial plane, 
a non-zero $B^\theta$ component that oscillates at the pulsar rotation rate $\Omega_*$. 
An Alfv\'en wavelength (propagating at $\sim c$) with $P = 6$ ms (case of PSR B1257+12) 
is of the order of 1800 km, which is less than a typical planetary radius. 
This would put the assumption of stationarity into question, though less severely
in the case of a standard pulsar with $P \sim 1$s, where the wavelength (about 300 000 km) would be much larger than the planetary radius.

In spite of the preliminary character of our model, 
our study shows that the consideration of Alfv\'en wings associated to planets orbiting pulsars deserve some attention. 

The question of the radio emissions possibly associated to the Alfv\'en wings will be addressed in a forthcoming paper. 
Such emissions would provide astronomers with observational data relevant to the wind/planet interaction.

\begin{acknowledgements}
The authors thank Silvano Bonazzola (LUTH, Obs. Paris-Meudon) for bringing us to the subject of this study. This research has made use of the SIMBAD database,
operated at CDS, Strasbourg, France, and The Extrasolar Planets Encyclopaedia 
(http://exoplanet.eu/index.php), maintained by Jean Schneider at the LUTH, and the SIO, at the Observatoire de Paris, France.

\end{acknowledgements}


\begin{figure}
\resizebox{\hsize}{!}{\includegraphics{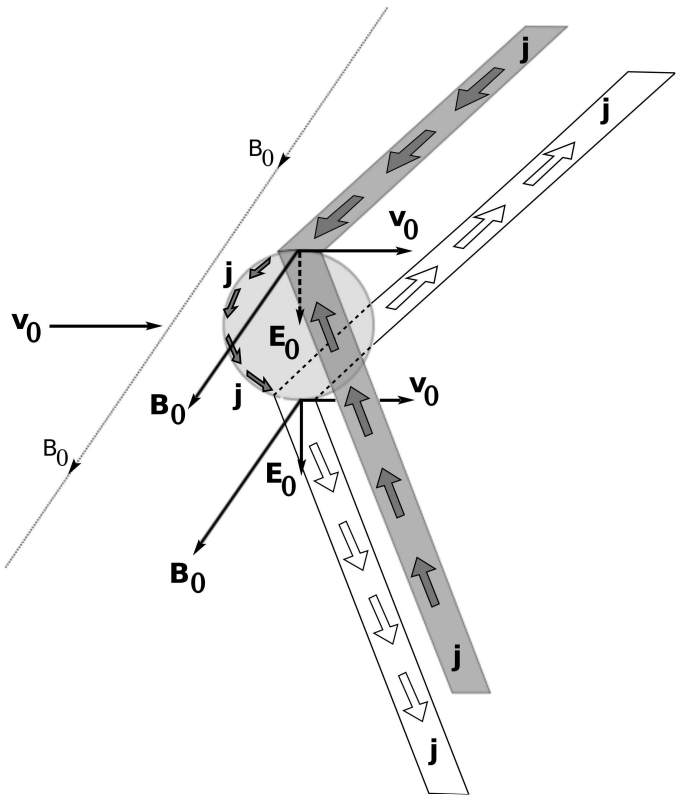}}
\caption{Schematic view of an unipolar inductor.}
\label{fig_inducteur_unipolaire}
\end{figure}

\begin{figure}
\resizebox{\hsize}{!}{\includegraphics{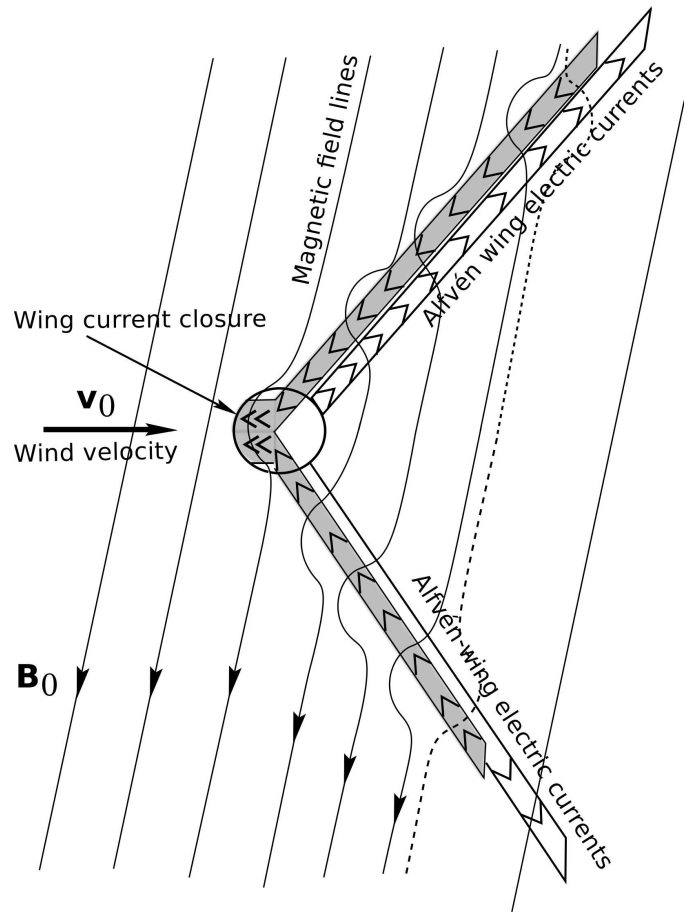}}
\caption{The Alfv\'enic wake of the planet seen from above the equatorial plane. 
}
\label{fig_ailealfperp}
\end{figure}

\end{document}